\documentclass[
 reprint,
 amsmath,
 amssymb,
 aps,
 longbibliography,
]{revtex4-1}

\usepackage[pdftex]{graphicx}
\usepackage{wasysym}
\usepackage{amsfonts}
\usepackage{ifsym}
\usepackage{todonotes}
\usepackage{color}
\usepackage{graphicx}
\usepackage{dcolumn}
\usepackage{bm}
\usepackage{hyperref}

\newcommand\net{%
\begin{tikzpicture}[scale=0.3\baselineskip/18pt]
 \draw (0,1) -- (1,1) -- (1,0) -- (2,0);
\end{tikzpicture}%
}

\newcommand{\nn}{\nonumber}

\newcommand{\COMMENT}[1]{}

\newcommand{\neqa}{\nonumber\end{eqnarray}}
\newcommand{\la}[1]{\label{#1}}

\renewcommand{\d}{\partial}

\newcommand{\<}{{\langle}}
\renewcommand{\>}{{\rangle}}

\newcommand{\re}{\relax{\rm I\kern-.18em R}}

\def\su2{{SU(2)}}

\def\[{\left[}
\def\]{\right]}

\def\({\left(}
\def\){\right)}
\def\[{\left[}
\def\]{\right]}

\def\<{\langle}
\def\>{\rangle}

\def\i2{\frac{i}{2}}

\def\cO{{\cal O}}

\def\2F1{\,_2{\rm F}_1}

\newcommand{\phaneq}{\phantom{{}=}}
\newcommand{\Li}{{\rm Li}}

\newcommand{\beq}{\begin{equation}}
\newcommand{\eeq}{\end{equation}}
\newcommand{\beqq}{\begin{equation*}}
\newcommand{\eeqq}{\end{equation*}}
\newcommand\beqa{\begin{eqnarray}}
\newcommand\eeqa{\end{eqnarray}}
\newcommand\beqaa{\begin{eqnarray*}}
\newcommand\eeqaa{\end{eqnarray*}}
\newcommand\bea{\begin{array}}
\newcommand\eea{\end{array}}

\begin{document}

\title{An Operator Product Expansion for Form Factors}

\author{Amit Sever$^{\displaystyle\footnotesize\net}$, Alexander G. Tumanov$^{\displaystyle\pentagon}$, Matthias Wilhelm$^{\displaystyle\Box}$}%

\vspace{15mm}

\affiliation{
$^{\displaystyle\footnotesize\net}$School of Physics and Astronomy, Tel Aviv University, Ramat Aviv 69978, Israel\\
$^{\displaystyle\pentagon}$Max-Planck-Institut f{\"u}r Physik, Werner-Heisenberg-Institut, 80805 M{\"u}nchen, Germany\\
$^{\displaystyle\Box}$Niels Bohr Institute, Copenhagen University, 2100 Copenhagen \O, Denmark
}%

\begin{abstract}
We propose an operator product expansion for planar form factors of local operators in \mbox{$\mathcal{N}=4$} SYM theory. This expansion is based on the dual conformal symmetry of these objects or, equivalently, the conformal symmetry of their dual description in terms of periodic Wilson loops. A form factor is decomposed into a sequence of known pentagon transitions and a new universal object that we call the ``form factor transition''. This transition is subject to a set of non-trivial bootstrap constraints, which are sufficient to fully determine it. We evaluate the form factor transition for MHV form factors of the chiral half of the stress tensor supermultiplet at leading order in perturbation theory and use it to produce OPE predictions at any loop order. We match the one-loop and two-loop predictions with data available in the literature. 
\end{abstract}

\maketitle

\section{Introduction}

The past ten years saw huge progress in our understanding of null polygonal Wilson loops, which was primarily motivated by the fact that these objects describe color-ordered scattering amplitudes in planar $\mathcal{N}=4$ SYM theory. Another motivation lies in them controlling a certain limit of correlation functions of local operators in this theory \cite{Alday:2010zy,Eden:2010zz,Eden:2010ce,Eden:2011yp,Eden:2011ku}. A further class of fundamental observables with a dual description in terms of certain null polygonal Wilson loops are form factors (FFs) \cite{Alday:2007he,Maldacena:2010kp,
Brandhuber:2010ad,Ben-Israel:2018ckc,Bianchi:2018rrj}; in terms of complexity, they lie somewhere in between scattering amplitudes and correlation functions.

The FF $\mathcal{F}_{\mathcal{O}}$ describes the overlap of a state created  by a local operator $\mathcal{O}$ with an $n$-particle asymptotic state:
\begin{align}
  \mathcal{F}_{\mathcal{O}}(k_1,\dots,k_n;q)=\!\int\! dx^4 e^{-ixq}\langle k_1,\dots,k_n|\mathcal{O}(x)|0\rangle
  \, ,
\end{align}
which has support on $q=\sum_i k_i$. While $k_i^2=0$, generically $q^2\neq0$. Because of the dependence on the local operator FFs are richer than scattering amplitudes, which themselves can be thought of as FFs of the identity operator. The simplest non-trivial operator to consider is the chiral half of the stress tensor supermultiplet, which contains the self-dual part of the Lagrangian, ${\cal F}_{\cal L}$. Operators in this multiplet preserve half of the supersymmetry  and their FFs can be classified according to the helicity of the external particles. In this letter, we focus on the simplest, maximally-helicity-violating (MHV) configuration of the color-ordered form factor, ${\cal F}_{\cal L}^\text{MHV}(k_1,\dots,k_n)$, which in many ways resembles MHV scattering amplitudes.

Many of the perturbative methods for computing scattering amplitudes have been generalized to FFs, see the recent review \cite{Yang:2019vag} for a detailed account. 
Moreover, integrable structures have been identified in FFs at strong coupling \cite{Maldacena:2010kp,Gao:2013dza} as well as at weak coupling \cite{Frassek:2015rka}.

At the non-perturbative level, the only systematic method of studying scattering amplitudes is the operator product expansion (OPE), which is based on dual conformal symmetry \cite{Drummond:2007au}. This powerful property of planar amplitudes is nothing but the conformal symmetry of their dual description in terms of null polygonal Wilson loops. The momenta of the particles, $k_i$, determine the positions of the cusps of this Wilson loop by the simple rule $x_{i+1}-x_i=k_i$. 
For ${\cal F}_{\cal L}^\text{MHV}$, the dual Wilson loop is determined in the same way. 
However, because the total momentum $q\ne0$, the corresponding contour is not closed, but periodic: $x_{i+n}-x_i=q$. The periodicity is also imposed at the quantum level and mixes the spacetime translation with the color trace \cite{Ben-Israel:2018ckc,Cavaglia:2020hdb}. As a result, this periodic Wilson loop is only defined in the planar limit. 
We will also refer to it as a {\it wrapped polygon}, since it is wrapped once around a cylinder topology. Similar to amplitudes, FFs are invariant under a version of dual conformal symmetry -- one that acts on both the cusps $x_i$ of the wrapped polygon and its periodicity constraint \cite{Ben-Israel:2018ckc,Chicherin:2018wes,Bianchi:2018rrj}.
The existence of this non-trivial symmetry suggests that the OPE method can be extended to FFs. In this letter, we present this extension explicitly.

The OPE is a decomposition of the FF into two types of universal building blocks. One of them is the pentagon transition, which is independent of the operator $\cO$ and has been bootstrapped at finite coupling in refs.\ \cite{Basso:2013vsa, Basso:2013aha,Basso:2014koa,Basso:2014nra,Basso:2014hfa,Basso:2015rta,Basso:2015uxa,Belitsky:2014sla,Belitsky:2014lta,Belitsky:2016vyq} using the integrability of the Gubser-Klebanov-Polyakov (GKP) flux tube. The other building block is the {\it form factor transition} that we introduce here, which encodes the information about $\cO$. We also expect that it can be bootstrapped at finite coupling.
One can draw an analogy between the FF OPE and the OPE for local operators in a CFT. The pentagon transitions are analogous to the three-point functions, while the FF transition is analogous to the one-point function that can arise at non-zero temperature or in the presence of a defect.

The remainder of this letter is organized as follows. In section \ref{secOPE} we describe the general features of the FF OPE. We define the FF transition and list the bootstrap constraints it has to satisfy in section \ref{secFF}. In section \ref{secBorn} we study the FF transition at leading order in perturbation theory, before we confront the OPE with the available perturbative data and make higher loop predictions in section \ref{seccheck}. Future direction are discussed in section \ref{secdisscussion}.

\section{Form Factor OPE}\la{secOPE}

\begin{figure}[h]
\centering
\includegraphics[width=7cm]{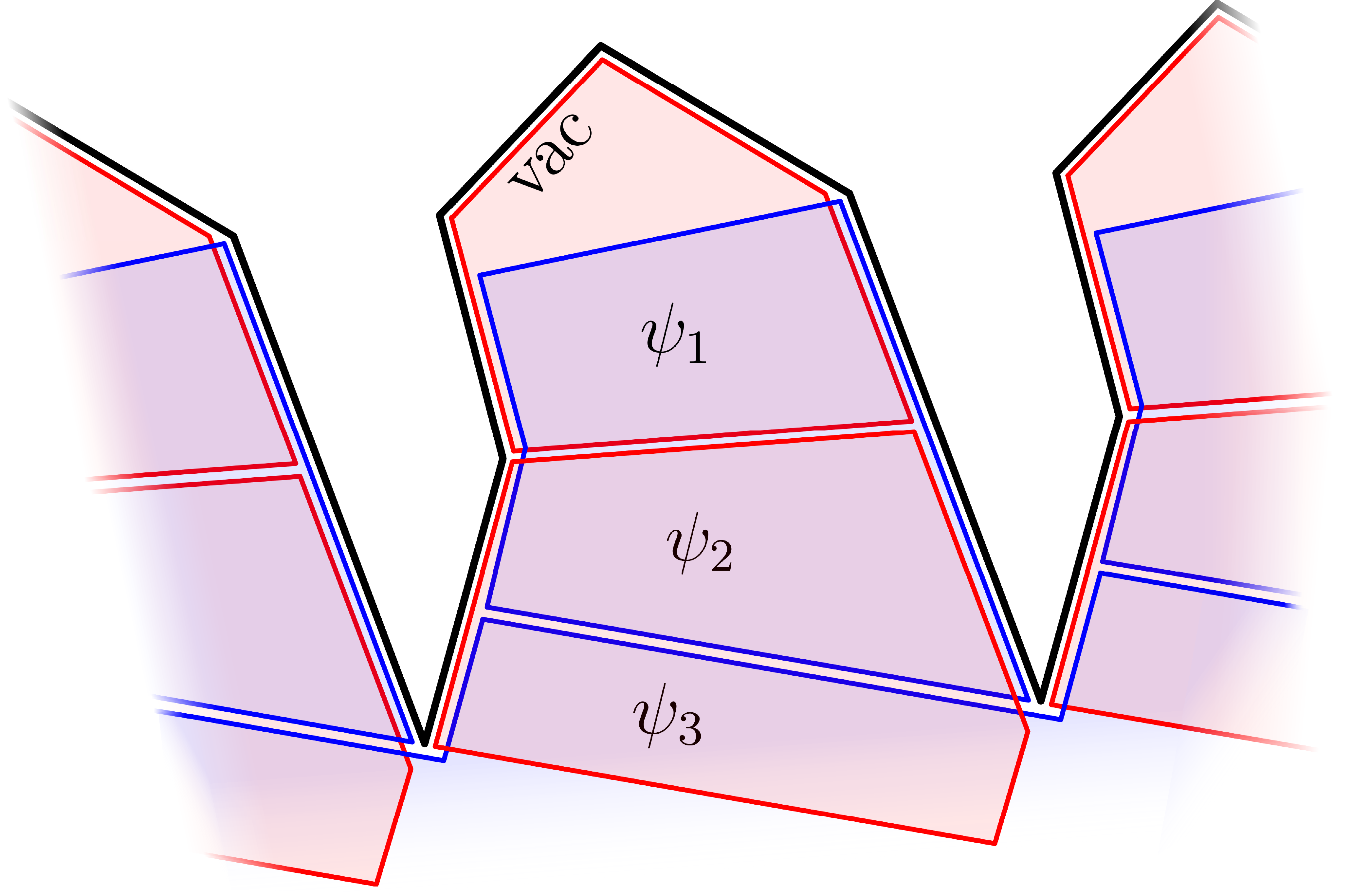}
\caption{Decomposition of an $n$-sided wrapped polygon into a sequence of pentagons and a two-sided wrapped polygon. Every two consecutive pentagons overlap on a null square and every two consecutive null squares form a pentagon. Similarly, the last pentagon overlaps with the two-sided wrapped polygon on a null square. Every square that arises from these overlaps shares two of its opposite cusps with the $n$-sided wrapped polygon. For the last (bottom) square, these two cusps coincide with one of the cusps of the two-sided wrapped polygon and its periodic image.
}
\label{OPEfig}
\end{figure}

We decompose an $n$-sided wrapped polygon into one two-sided wrapped polygon and $n-2$ pentagons, which overlap on $n-2$ ``middle squares", as shown in figure~\ref{OPEfig}. 

\begin{figure}[b]
\centering
\includegraphics[width=8.5cm]{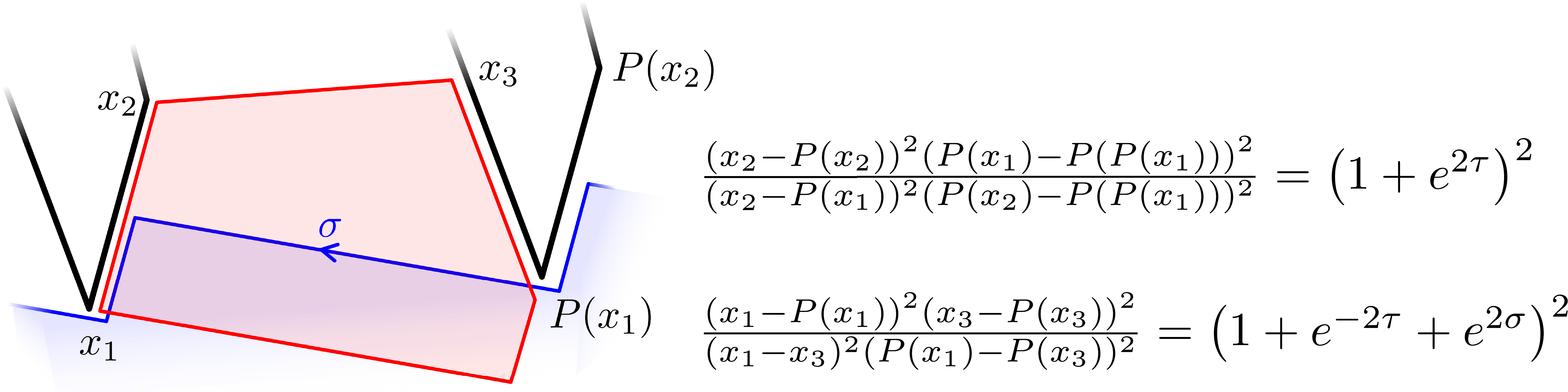}
\caption{\small We associate two independent conformal cross ratios, $\tau$ and $\sigma$, to the last OPE channel. For an $n$-particle FF, these are denoted as $\tau_{n-2}$ and $\sigma_{n-2}$. Here, $P(x)$ stands for the periodic image of the point $x$. Under a conformal transformation $x\to K(x)$, the periodic image transforms as $P(x)\to\tilde P(K(x))=K(P(x))$. In a conformal frame in which $P$ is a translation, $P(x)-x$ is independent of the point $x$ and the expressions for the two conformal cross ratios in the figure simplify to complete squares. 
}
\label{geometryfig}
\end{figure}
A null square is invariant under three commuting conformal symmetries; they are parametrized by $\tau$, $\sigma$ and $\phi$ that are conjugate to the GKP twist, conformal spin and angular momentum in the transverse plane, respectively. We use these symmetries to parametrize all conformally inequivalent $n$-sided wrapped polygons by a set of $3n-7$ independent conformal cross ratios as follows. To squares that are associated with an overlap of two pentagons, we assign three conformal cross ratios $\{\tau_i,\sigma_i,\phi_i\}_{i=1}^{n-3}$ that are defined in the same way as for closed polygons, see figure 2 in ref.\ \cite{Basso:2013vsa}. Geometrically, shifting these variables amounts to acting with the conformal symmetries of that square on all cusps above it; cf.\ figure \ref{OPEfig}. Similarly, for the last square the conformal cross ratios $\tau_{n-2}$ and $\sigma_{n-2}$ are defined in figure \ref{geometryfig}, while $\phi_{n-2}=0$ \footnote{For two overlapping pentagons, $\phi_i$ is the angle between the two opposite cusps in the plane that is transverse to the corresponding square. For the overlap of a pentagon with the two-sided wrapped polygon, only one cusp and its periodic images lay outside of the plane of the corresponding square; since the periodicity constraint acts within the plane of the square, their relative angle in the transverse plane vanishes.
}.

\begin{figure}[t]
\centering
\includegraphics[width=8.5cm]{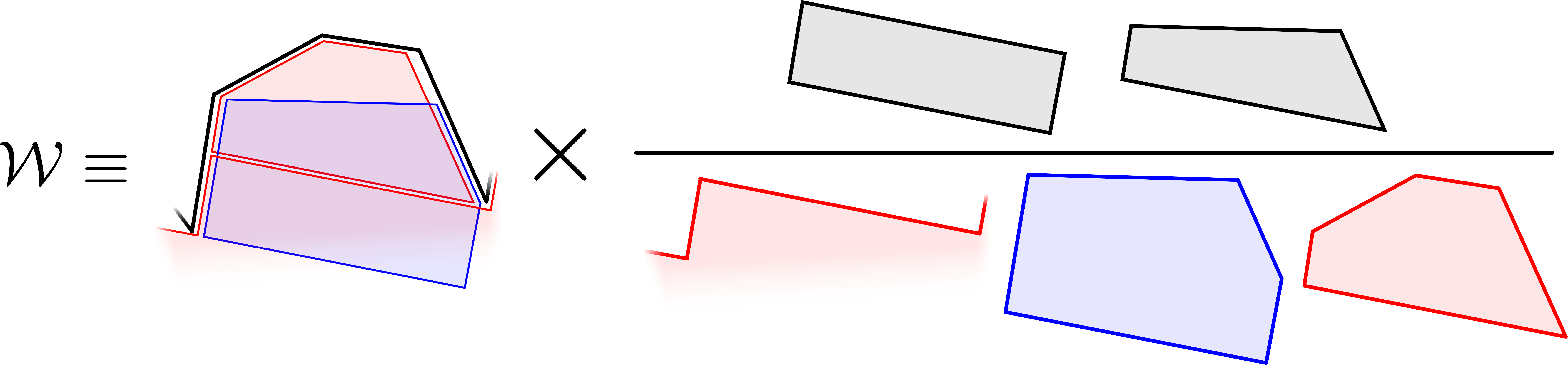}
\caption{A finite conformally invariant ratio is constructed by multiplying the $n$-sided wrapped polygon by all squares except for the first and dividing by all pentagons as well as the two-sided wrapped polygon, 
$
\mathcal{W}_{n}= \frac{\langle W_{n\text{-pt ff}}\rangle\times\langle W_\text{2nd square}\rangle\langle W_\text{3rd square}\rangle\ldots}{\langle W_{2\text{-pt ff}}\rangle\times\langle W_\text{1st pentagon}\rangle\,\langle W_\text{2nd pentagon}\rangle\,\ldots}
$.
Here, this ratio is illustrated for $n=4$.
}
\label{ratiofig}
\end{figure}

The UV divergences of the periodic Wilson loops are regularized using the pentagons, the squares and the two-sided wrapped polygon of the decomposition. Concretely, we are considering the ratio $\mathcal{W}_n$ defined in figure \ref{ratiofig}. 

The operator product expansion is the large $\tau$ expansion in which the sides of the polygon on top of each square 
are decomposed into a superposition of GKP flux-tube excitations. This flux is sourced by two opposite sides of the corresponding square. 
We start with the vacuum state in the 
top square in figure \ref{OPEfig}. It undergoes a series of pentagon transitions from one square to the next, with an eigenstate in the $i^{\text{th}}$ channel denoted as $\psi_i$. The propagation of this state 
results in the factor $\exp(-E_i\tau_i+ip_i\sigma_i+im_i\phi_i)$, where $\{E_i,p_i,m_i\}$ are the GKP energy, momentum and angular momentum, respectively. Finally, the state $\psi_{n-2}$ is absorbed by the two-sided periodic Wilson loop. 
We call this final step the {\it form factor transition}. In summary, this sequence of transitions and propagation 
can be written as
\begin{equation}\la{OPEdecomp}
\begin{aligned}
 \mathcal{W}_{n}&=\sum_{\psi_1,\dots,\psi_{n-2}}e^{\sum_j(-E_j\tau_j+ip_j\sigma_j+i m_j\phi_j)}\\
 &\phaneq\times \mathcal{P}(0|\psi_1)\dots\mathcal{P}(\psi_{n-3}|\psi_{n-2})\,\mathcal{F}(\psi_{n-2})\ .
 \end{aligned}
\end{equation}
Here, $\mathcal{P}$ denotes the pentagon transition and 
$\mathcal{F}$ is the form factor transition. 

The decomposition \eqref{OPEdecomp} applies to periodic Wilson loops in any conformal theory with a stable flux between fast-moving quarks. For the rest of this letter we focus on ${\cal N}=4$ SYM theory, in which periodic Wilson loops are dual to form factors. Under this duality, the OPE maps to the expansion around the multi-collinear limit. Moreover, the GKP flux-tube dynamics of this theory is integrable. 
Therefore, we expect to be able to bootstrap the building blocks entering eq.\ \eqref{OPEdecomp} at finite 't~Hooft coupling.

The basis of GKP eigenstates as well as their dispersion relations have been constructed in ref.\ \cite{Basso:2010in}. The pentagon transitions and integration (or square) measures have been bootstrapped in refs.\ \cite{Basso:2013vsa, Basso:2013aha,Basso:2014koa,Basso:2014nra,Basso:2014hfa,Basso:2015rta,Basso:2015uxa,Belitsky:2014sla,Belitsky:2014lta,Belitsky:2016vyq}. Hence, in order to compute planar form factors in ${\cal N}=4$ SYM theory, all that remains is to bootstrap one new building block -- the form factor transition. This object is universal; it does not depend on the number of particles or their kinematical configuration, but only on the local operator and the GKP eigenstate. In the next sections, we study the FF transitions for the chiral part of the stress tensor supermultiplet. Before, let us set our notations, which are aligned with the ones introduced in ref.\ \cite{Basso:2013vsa}.

The simplest form factor $\mathcal{F}_{\mathcal{L}}$ that admits a nontrivial OPE decomposition is the three-point MHV one. For this case, we have
\beq\la{W3}
\mathcal{W}_3= \sum\limits_{\bf a}\int d\textbf{u}\,P_{\bf a}(0|\textbf{u})\,F_{\bar{{\bf a}}}(\bar{\textbf{u}})\,e^{-\tau E(\textbf{u})\,+\,i\sigma p(\textbf{u})}\ ,
\eeq
where we sum over the complete basis of GKP eigenstates. These states are parametrized by the number of excitations $N$, their species ${\bf a}=\{a_1,\dots,a_N\}$ and their flux-tube momenta or, equivalently, their Bethe rapidities ${\bf u}=\{u_1,\ldots,u_N\}$, with $\bar{\bf a} = \{a_N,\ldots,a_1\}$ and $\bar{\textbf{u}}=\{-u_N,\dots,-u_1\}$. Here, $P_{\bf a}$ are the pentagon transitions, and the integration measure is given by
\beq\label{Na}
d\textbf{u} = \mathcal{N}_{\bf a}\,\prod\limits_{i=1}^N \mu_{a_i}(u_i)\,\frac{du_i}{2\pi}\ ,
\eeq
with $\mu_a$ being the single-particle measures and ${\cal N}_{\bf a}$ being a symmetry factor. Lastly, $F_{\bf a}$ are the FF transitions that will be studied in the following two sections.

\section{The Form Factor Transition}\la{secFF}

\begin{figure}[h]
\centering
\includegraphics[width=7.5cm]{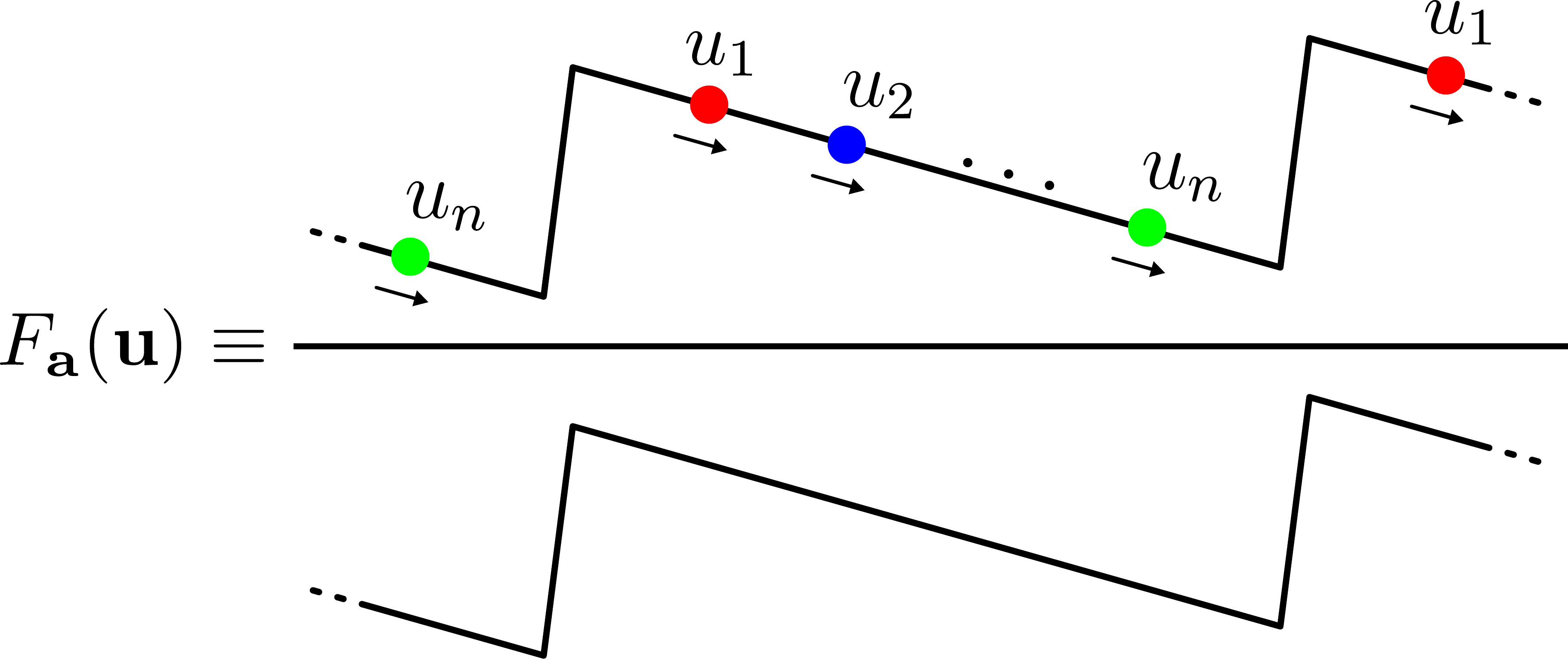}
\caption{The FF transition 
is given by the ratio between the expectation value of the two-sided wrapped polygon with and without GKP excitations inserted on its base. 
}
\label{FFtransdefinition}
\end{figure}

The FF transition computes the amplitude for a GKP in-state to be absorbed by the two-sided wrapped polygon, see figure \ref{FFtransdefinition}. 
It is subject to a set of constraints that we list below. These constraints are similar to those obeyed by integrable two-dimensional form factors of a branch-point operator of angle $\pi$ 
\footnote{For relativistic theories, the Watson's equations are spelled out in ref.\ \cite{Cardy:2007mb}. The FF transition corresponds to the case $k=\frac{1}{2}$ with a modified factorization pole.}.
In ref.\ \cite{Toappear2}, we use them to bootstrap the FF transitions at finite 't~Hooft coupling.

{\it Watson --} Reordering two adjacent excitations within a state is equivalent to acting on it with the S-matrix. This property is inherited by the FF transition:
\beq\la{Watson}
F(...,u_j,u_{j+1},...)=S(u_j,u_{j+1})F(...,u_{j+1},u_{j},...)\ ,
\eeq
where we have suppressed the species index.

{\it Singlet --} The two-sided wrapped polygon is invariant under a 
$U(1)_\phi\times SU(4)_R$ symmetry, where  
the $U(1)_\phi$ factor corresponds to rotations in the two-dimensional transverse plane and $SU(4)_R$ is the R-symmetry group. As a result, the FF transition must be a $U(1)_\phi\times SU(4)_R$ singlet:
\beq
F_{a_1,\dots,a_n}({\bf u})={\cal M}_{a_1}^{b_1}\dots{\cal M}_{a_n}^{b_n}\,F_{b_1,\dots,b_n}({\bf u})\ ,
\eeq
where ${\cal M}\in U(1)_\phi\times SU(4)_R$. As such, it can only absorb singlet states.

As the fundamental GKP excitations are all charged under $U(1)_\phi\times SU(4)_R$, the FF transition cannot absorb a single-particle excitation. 
Moreover, 
only singlet states with even Born-level energy can contribute to the FF transition \footnote{One can see that $m_i$, the $U(1)_\phi$ charge of the $i^{\text{th}}$ excitation, $t_i$, the number of Gra\ss{}mann variables $\theta_i$ associated with it, and $E_i$, its Born-level energy, are related by $(m_i + t_i/2)\,{\rm mod}\,2 = E_i\,{\rm mod}\,2$. Combining this with the singlet constraints, $\sum_i m_i=0$ and $\sum_i t_i\,{\rm mod}\,4 = 0$, leads to $\sum_i E_i\,{\rm mod}\,2 = \sum_i (m_i + t_i/2)\,{\rm mod}\,2 = (\sum_i m_i) + (\sum_i t_i)/2\,{\rm mod}\,2 = 0$.}. As a result, at any loop order only even powers of $e^{-\tau}$ can appear in the large $\tau$ expansion \eqref{W3}.

{\it Reflection --} In addition to the continuous symmetries above, the two-sided wrapped polygon is also invariant under a discrete ${\mathbb Z}_2$ symmetry. It acts by flipping the direction of the two edges. This transformation has the effect of inverting the $\sigma$ direction. As a result, 
the FF transition is subject to the relation \footnote{The OPE tessellation is alternating and therefore not invariant under this reflection. As for the scattering amplitude OPE, this is only reflected in the $i\epsilon$ prescription for the 
fermions' measure, see the discussion in ref.\ \cite{Basso:2014koa}.}
\beq\la{reflection}
F_{\bf a}({\bf u})=F_{\bar{\bf a}}(\bar{\bf u})\ .
\eeq

{\it Square limit --} The FF transition and the measure are related by 
\begin{align}\la{squarelimit}
\lim_{u_1\to u_n} 
F_{\bf a}({\bf u})&=\frac{-\,i\delta_{a_n,\bar a_1}}{\mu_{a_1}(u_1)}\frac{F_{a_2,\dots,a_{n-1}}(u_2,\dots,u_{n-1})}{ u_1-u_n-i\epsilon}\\
&\pm\Big(S(u_1,u_n)\prod\limits_{1<j<n}S(u_1,u_j)S(u_j,u_n)\Big)_{\bf a}^{\bf b}\nn\\
&\times\frac{-\,i\delta_{b_n,\bar b_1}}{\mu_{b_1}(u_1)}\frac{F_{b_2,\dots,b_{n-1}}(u_2,\dots,u_{n-1})}{ u_n-u_1-i\epsilon}\ , \nn
\end{align}
where the plus sign is for bosons and the minus sign for fermions. 

This relation can be understood as follows. In position space, the
residue at $u_1=u_n$ controls the regime of the FF transition where  $|\sigma_1-\sigma_n|\to\infty$ with $\sigma_1+\sigma_n$ and $\sigma_{j}$ for $1<j<n$ fixed. For $\sigma_n>\sigma_1$ ($\sigma_n<\sigma_1$), this limit corresponds to sending the first excitation towards the left (right) edge and the $n^\text{th}$ excitation towards the right (left) edge in figure \ref{FFtransdefinition}. Since the spectrum is gapped and the right edge is the periodic image of the left edge, the two excitations decouple and propagate as if they where inserted on the top and the bottom of a square. The first and second term in eq.\ \eqref{squarelimit} correspond to the limit in which the first excitation is sent towards the left and right edge, respectively. 
To reach the second limit, the first and the last excitations have to pass through all other excitations as well as through each other. According to eq.\ \eqref{Watson}, this results in the product of S-matrices appearing in the second term in eq.\ \eqref{squarelimit}. 

\begin{figure}[h]
\centering
\includegraphics[width=7.5cm]{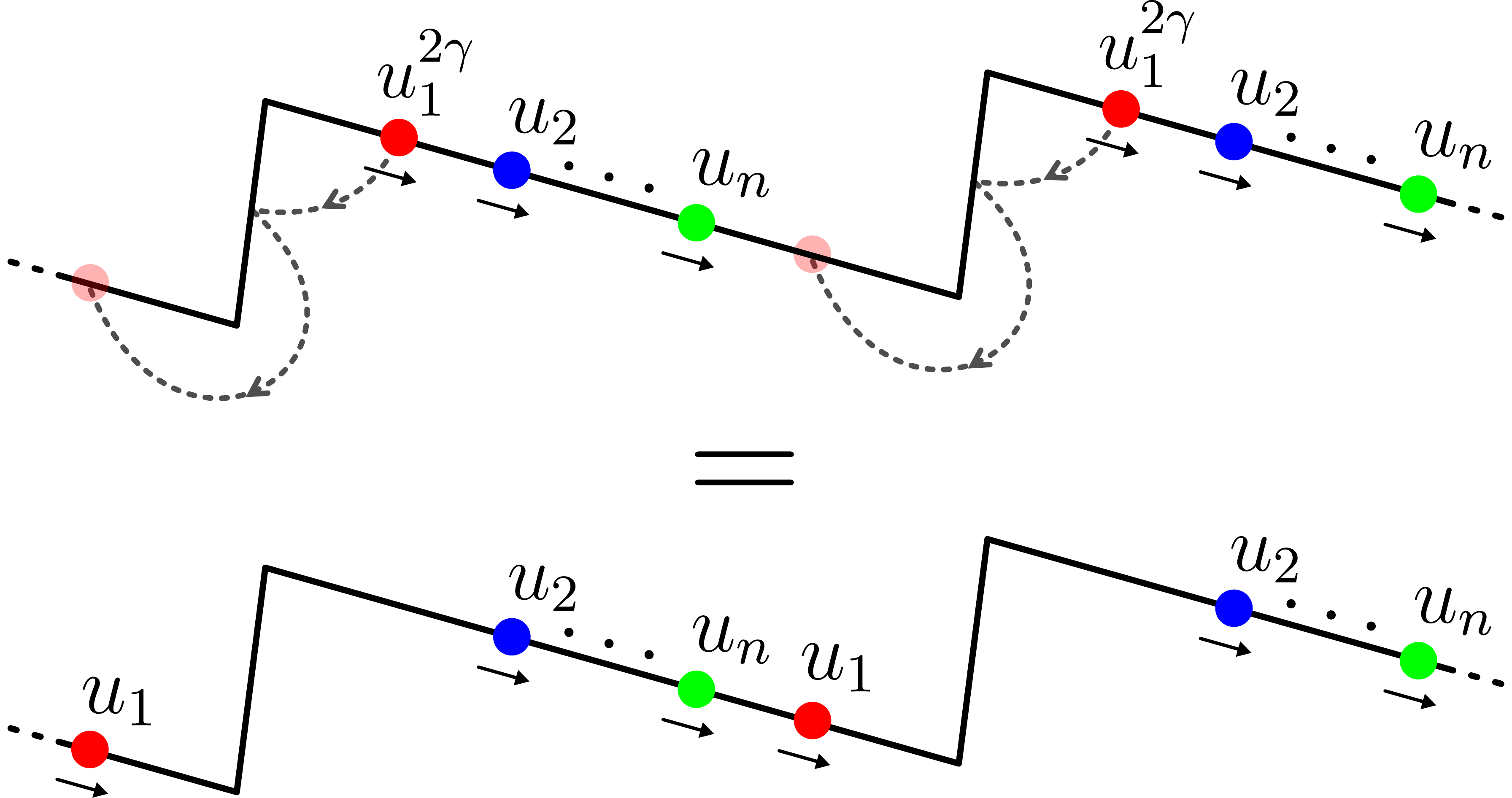}
\caption{Applying a mirror transformation to the first excitation is equivalent to transporting it to the neighboring edge on the left. After two successive mirror transformations, or a crossing transformation, the first excitation becomes the last one.}\la{crossingfig}
\end{figure}
{\it Crossing --} The most non-trivial constraint has to do with the crossing symmetry of the transition and is depicted in figure \ref{crossingfig}. It reads
\beq
F(u_1^{2\gamma},u_2,\dots,u_n)=F(u_2,\dots,u_n,u_1)\ .
\eeq
Here, $u^\gamma$ is a mirror transformation such that $p(u^\gamma)=iE(u)$ and $E (u^\gamma)=ip(u)$, see ref.\ \cite{Basso:2010in}.

\section{FF transitions at Born level}\la{secBorn}

The leading contribution to the OPE \eqref{W3} comes from the lightest singlet state. In perturbation theory, this contribution stems from three two-particle singlet states and two one-particle effective excitations, all of which have the same tree-level energy $E=2$. Each of the three two-particle singlet states is a superposition of all possible singlet combinations of two scalar ($\phi\bar\phi$), two fermion ($\psi\bar\psi$) and two gluon ($F\bar{F}$) fields inserted 
on the base of the wrapped polygon in figure \ref{FFtransdefinition}. These states differ in the asymptotic limit, in which the two fields are taken far apart. Only one out of the three pairs of fields survives in this limit, and this is the pair that labels the state. The two effective one-particle excitations $F_{+-}\pm F_{z\bar{z}}$ correspond to parts of the two-fermion singlet state that behave as independent single-particle excitations at weak coupling. We discuss them at the end of this section.

In ref.\ \cite{Sever:2021nsq}, 
we have explicitly constructed the aforementioned superpositions that correspond to the three two-particle singlet states at Born level. We will now use them to compute the Born-level FF transitions we denote by $F_{\phi\bar\phi}$, $F_{\psi\bar\psi}$ and $F_{F\bar F}$. 

At leading order in perturbation theory, the expectation value of the two-sided wrapped polygon is equal to 1. Hence, only the numerator in figure \ref{FFtransdefinition} contributes to the transition non-trivially. Consider the wrapped polygon with two conjugate fields inserted at positions $\sigma_1$ and $\sigma_2$ with $\sigma_2>\sigma_1$. At Born level, we obtain the propagator between the field at $\sigma_2$ and the periodic image of the field at $\sigma_1$:
\beq\la{FFtree}
\includegraphics[width=7.5cm]{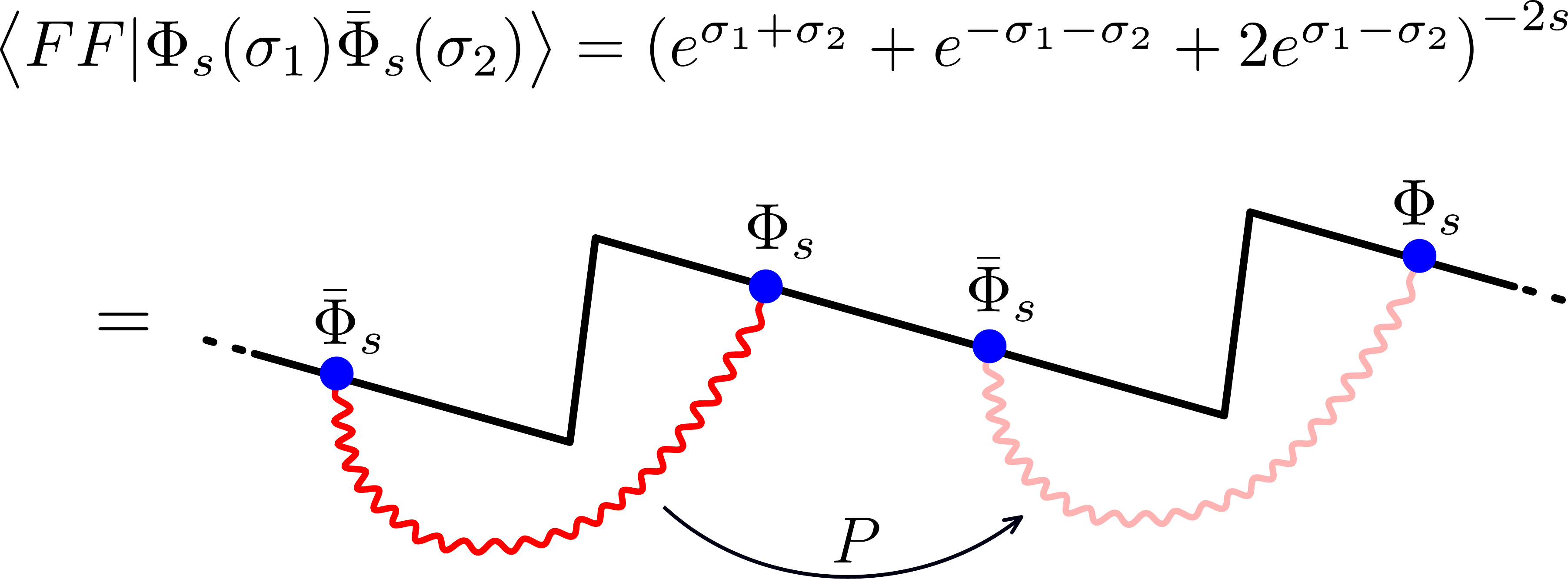}
\eeq
where $\Phi_s$ is a field of conformal spin $s=\frac{1}{2},1,\frac{3}{2}$ for scalars, fermions and gluons respectively.

Convoluting the singlet states given in ref.\ \cite{Sever:2021nsq} with the propagator \eqref{FFtree}, we arrive at 
\begin{align}\label{BornlevelFF}
&F_{\phi\bar\phi}(u,v)= -\,6\times\frac{4}{g^2\left(u-v-2i\right)\left(u-v-i\right)}\nn\\ &\hphantom{F_{\phi\bar\phi}(u,v)=-\,3}\times\frac{\Gamma\left(iu-iv\right)}{\Gamma\left(\frac{1}{2}+iu\right)\Gamma\left(\frac{1}{2}-iv\right)}\,,\\\nn
&F_{\psi\bar\psi}(u,v)= +\,4\times\frac{2}{g^2}\,u\sinh(\pi u)\,\delta(u-v)\ ,\\
&F_{F\bar{F}}(u,v)= -\,1\times \frac{2}{g^2}\left(u^2+\tfrac{1}{4}\right)\cosh(\pi u)\,\delta(u-v)\ ,\nn
\end{align}
where $g^2=\frac{g_{YM}^2N}{16\pi^2}$. Factors of 6, 4 and 1 for the number of real scalars, fermions and gluons are included here for convenience, such that no further summation over the flavor index is required in eq.\ \eqref{W3}. Note that for two gluons and two fermions, the right-hand side of eq.\ \eqref{squarelimit} reduces to a delta function divided by the measure. We see that for these states the full Born level result is given solely by this simple square limit contribution. This might be surprising, because the contribution of each type of fields to eq.\ \eqref{BornlevelFF} is highly non-trivial, but they combine to an almost trivial result. 

Lastly, we consider the two one-particle effective excitations, which consist of the fields $F_{+-}\pm F_{z\bar{z}}$ inserted on the base of the wrapped polygon in figure~\ref{FFtransdefinition}. There are two types of diagrams that contribute to the corresponding FF transitions at Born level. One consists of a cubic interaction vertex periodically contracted with itself and with the field insertion. We find that the contribution of this type vanishes due to the cancellation between the different types of fields running in the loop. The other type is the periodic contraction of the commutator $[A_+,A_-]$. 
The two orderings cancel each other, leading to the corresponding transitions vanishing in total.

\section{Perturbative Tests and Predictions}\la{seccheck}

We now perform a perturbative test of the FF OPE and use it to make higher loop predictions. 

We start by extracting the OPE data from previously computed form factors \cite{Brandhuber:2010ad}. At one-loop order, we find the ratio ${\cal W}_{n=3}$ defined in figure \ref{ratiofig} to be given by
\begin{align}\label{RRU1}
\mathcal{W}_{3}^{(1)} &=4\sigma^2 - 2\,\Li_2(-\,e^{-2\tau}) + 2\,\Li_2(-\,e^{-2\tau} - e^{2\sigma})\nonumber\\
&\phaneq+ 2\,\Li_2(-\,e^{-2\tau} - e^{-2\sigma}(1 + e^{-2\tau})^2)+\frac{\pi^2}{3}\ ,
\end{align}
where $\mathcal{W}_{3}=1+\sum_{\ell=1}^\infty g^{2\ell}\mathcal{W}_{3}^{(\ell)}$.
As expected from the singlet axiom, the large $\tau$ expansion of $\mathcal{W}_{3}^{(1)}$ contains only even powers of $e^{-\tau}$, with the leading one given by
\begin{align}\label{TT}
\mathcal{W}_{3}^{(1)}&=2\,e^{-2\tau}\left(1 - 2\,\sigma\,e^{-2\sigma} - 4\cosh^2(\sigma)\log\left(1+e^{-2\sigma}\right)\right)\nonumber\\
&\phaneq+O(e^{-4\tau})\ .
\end{align}

On the OPE side, we insert the Born-level FF transitions \eqref{BornlevelFF} into eq.\ \eqref{W3} and perform the integration over the two rapidities, finding a perfect match with eq.\ \eqref{TT} \footnote{Note that since the scalars are real the corresponding $\mathcal{N}_{\bf a}$ factor from eq. (\ref{Na}) is equal to $\frac{1}{2}$.}.

Even without the higher loop FF transitions, we can already make certain all-loops predictions. Namely, at $\ell$-loop order we can predict the term with the highest power of $\tau$, i.e. $\tau^{\ell-1}e^{-2\tau}$. It is given by pulling down $(\ell-1)$ powers of the one-loop correction to the energy $g^2(E_{2s}^{(1)}(u_1)+E_{2s}^{(1)}(u_2))$ from the exponent $e^{-\tau(E_{2s}(u_1)+E_{2s}(u_2))}$. The one-loop correction to the energy of the individual excitations is given by  \cite{Basso:2010in}  
\beq
E_{2s}^{(1)}(u) =2\[\psi(s+iu)+\psi(s-iu)- 2\psi(1)\]\ ,
\eeq
where $\psi(x)=\frac{\Gamma'(x)}{\Gamma(x)}$ is Euler's digamma function.

With the two-loop data available for the three-point form factor reminder function ${\cal R}_3$  \cite{Brandhuber:2012vm}, we can test the OPE prediction for the $\tau e^{-2\tau}$ term in ${\cal W}_3$ at two-loop order.
To do so, one first has to translate between these two finite dual conformally invariant functions, ${\cal R}$ and ${\cal W}$. They are related as
\beq\la{RtoW}
{\cal W}_n=\exp\left[\frac{\Gamma_{\text{cusp}}}{4}\,\mathcal{W}^{(1)}_{n}\right]\times \mathcal{R}_n\ ,
\eeq
where $\Gamma_{\text{cusp}}=4g^2+\dots$ is the cusp anomalous dimension.
Using eq.\ \eqref{RtoW}, we obtain the following result:
\begin{equation}\label{TTlogT}
 \begin{aligned}
\mathcal{W}^{(2)}_{3,\,\tau e^{-2\tau}}=&-8\left[1 - \left(1 + e^{-2\sigma}\right)\log\left(1+e^{2\sigma}\right)\right]\\
&\ \ \times\left[1 - \left(1 + e^{2\sigma}\right)\log\left(1+e^{-2\sigma}\right)\right] ,
\end{aligned}
\end{equation}
which is in perfect agreement with the OPE prediction.

At three-loop order, we can predict the term proportional to $\tau^2e^{-2\tau}$:
\begin{align}\label{tausq3}
&\mathcal{W}^{(3)}_{3,\,\tau^2e^{-2\tau}}=\\
&\frac{16}{3}\cosh^2(\sigma)\log\left(1+e^{-2\sigma}\right)\Big[12\sigma\left(4-3\sigma\right)-24-\pi^2\nn\\
&+8\log\left(1+e^{-2\sigma}\right)\left(3-6\sigma-2\log\left(1+e^{-2\sigma}\right)\right)\Big]+\frac{4\pi^2}{3}\nn\\
&+24-16\sigma\left(4-3\sigma+4e^{-2\sigma}\right)- 32\cosh^2(\sigma)\,\Li_3\left(-\,e^{-2\sigma}\right) .\nn
\end{align}
Similarly, we can produce higher loop predictions; we refrain from giving these explicitly due to their size.

\section{Discussion}\la{secdisscussion}

In this letter, we have introduced an operator product expansion for form factors in planar ${\cal N}=4$ SYM theory. It reduces the computation of the dual periodic Wilson loop to known fundamental building blocks \cite{Basso:2013vsa,Basso:2014koa} and a single new universal building block -- the FF transition.  

We have calculated the two-particle FF transition at Born level \eqref{BornlevelFF}. A natural finite-coupling ansatz for the gluonic and fermionic two-particle FF transitions that is consistent with all the constraints is  \footnote{To analytically continue the delta function along the crossing path, we can express it as a pair of poles as was done in eq.\ \eqref{squarelimit}.}
\begin{align}\la{gluonfermiontrans}
F_{\Phi\bar\Phi}(u,v) \propto \frac{\delta(p_\Phi(u)-p_\Phi(v))}{\mu_\Phi(u)}\times\frac{\d p_\Phi(u)}{\d u}\ ,
\end{align}
where $\Phi\in\{\psi,F\}$ and $p_\Phi(u)$ is the GKP momentum. Based on this conjecture, the bootstrap constraints for the FF transition that we formulated in this letter, and the perturbative data available to us, we were able to fix the remaining scalar two-particle FF transition at finite coupling \cite{Toappear2}.
Transitions involving more than two flux-tube excitations can hopefully be fixed in terms of the two-particle ones using integrability. Our construction therefore opens the door for finite-coupling computations of FFs.

There are multiple future directions to pursue, some of which we list below. I) At strong coupling, the FFs are computed by minimizing the area of a periodic string in $AdS_5$ \cite{Maldacena:2010kp, Gao:2013dza}. We expect the corresponding Yang-Yang functional to be constructed from the gluon and fermion FF transitions \eqref{gluonfermiontrans} along with the corresponding pentagon transitions and measures. II) In this letter, we have only considered MHV FFs of the chiral part of the stress tensor supermultiplet. It would be interesting to extend our considerations to the N$^k$MHV case, for which the result is expected to be given by a version of the super-periodic Wilson loop introduced in ref.\ \cite{Ben-Israel:2018ckc}. In parallel, it would be interesting to bootstrap the corresponding charged FF transitions, in analogy to the charged pentagon transitions of refs.\  \cite{Basso:2014hfa,Basso:2015rta}. III) Another interesting direction is to consider local operators other than the chiral part of the stress tensor supermultiplet; corresponding FFs have been studied in refs.\ \cite{Engelund:2012re,Brandhuber:2014ica,Wilhelm:2014qua, Nandan:2014oga, Loebbert:2015ova, Brandhuber:2016fni, Loebbert:2016xkw, Caron-Huot:2016cwu,Banerjee:2016kri,Ahmed:2016vgl,Brandhuber:2018xzk}. T-duality is expected to map their higher integrability Yangian charges into dual ones \cite{Beisert:2008iq,Beisert:2009cs} that are evaluated along one period of the dual Wilson loop. IV) It is possible to extend the hexagon function program of refs.\ \cite{Dixon:2011pw,Dixon:2013eka,Dixon:2014iba,Caron-Huot:2016owq} to analogous FF functions \cite{Dixon:2020bbt}. The interplay between the OPE and these FF functions provides a plethora of valuable checks of our predictions and vice versa. V) Finally, it would be interesting to see if our considerations can be used for studying FFs in other theories like ABJM \cite{Aharony:2008ug}.

{\it Acknowledgements:} 
We are very grateful to B.~Basso for many valuable discussions and comments on the draft. AT and MW are grateful to CERN for hospitality. AS is grateful to NBI for hospitality. AS was supported by the I-CORE Program of the Planning and Budgeting Committee, The Israel Science Foundation (grant number 1937/12) and by the Israel Science Foundation (grant number 1197/20). AT received funding from the European Research Council (ERC) under the European Unions Horizon 2020 research and innovation programme, Novel structures in scattering amplitudes (grant agreement No. 725110).
MW was supported in part by the ERC starting grant 757978 and the research grants 00015369 and 00025445 from Villum Fonden.

\bibliography{bib}
\end{document}